\documentclass{PoS}

\title{New Tests of NRQCD from Quarkonia Within Jets}

\ShortTitle{Quarkonia in Jets}

\author{\speaker{Thomas Mehen} 
\\
        Duke University\\
        E-mail: \email{mehen@phy.duke.edu}}

\abstract{I review the current status of quarkonium production theory based on the non-relativistic QCD factorization formalism (NRQCD). While this theory describes much of the world's data on $J/\psi$ and $\Upsilon$ production, there are still outstanding problems, most notably the polarization of quarkonia at large $p_T$ in hadron colliders. In this talk we will present new tests of NRQCD involving the distribution of quarkonia within jets. The distribution of hadrons within jets is determined by nonperturbative functions called fragmenting jet functions (FJFs). FJFs are convolutions of fragmentation functions, evolved to the jet energy scale, with perturbatively calculable matching coefficients. I show how the FJFs for quarkonia can be calculated in NRQCD in terms of a few NRQCD long-distance matrix elements (LDME), so the dependence of the cross section on the energy fraction of the heavy quarkonium, $z$, is sensitive to the underlying production mechanism, and therefore provides a new test of NRQCD.  The jet energy and $z$ dependence of the cross section can be used to perform an independent extraction of NRQCD LDME. Finally, I describe ongoing work building on this result. This includes comparison of analytic resummed calculations with Monte Carlo simulations for two-jet events in $e^+e^-$ collisions with $B$ mesons, and three-jet events with $J/\psi$, as well as the definition of boost invariant jet substructure variables and calculation of a boost invariant soft function that are necessary for analytic calculations of jet cross sections at the Large Hadron Collider. }

\FullConference{QCD Evolution 2015 -QCDEV2015-\\
		26-30 May 2015\\
		Jefferson Lab (JLAB), Newport News Virginia, USA}

\begin{document}

\section{Status of Quarkonium Production Theory}

The most widely used approach to calculating the production cross section for quarkonia in hadronic collisions is the NRQCD factorization formalism~\cite{Bodwin:1994jh}. 
The production cross section for a quarkonium is written as the sum over products of short distance cross sections for producing the heavy quark-antiquark pair, $Q\bar{Q}$,
in a given angular momentum and color state multiplied by NRQCD long-distance matrix elements (LDME) which describe the hadronization of the $Q\bar{Q}$ into a final state including the quarkonium. Because the cross section for producing the $Q\bar{Q}$ is characterized by the large scale $2 m_Q$ or possibly other large scales such as the transverse momentum, $p_T$,  this part of the factorization theorem is calculable as a power series in $\alpha_s$. The transition of the $Q\bar{Q}$ into the quarkonium is necessarily non-perturbative, but this involves physics governed by the long-distance scales $m_Q v$, $m_Q v^2$, and $\Lambda_{\rm QCD}$, where $v$ is the typical relative velocity of the $Q\bar{Q}$ within the quarkonium. This transition is governed by the QCD multipole expansion and the  LDME scale with powers of $v$ according to the velocity scaling rules of NRQCD~\cite{Bodwin:1994jh}. Thus in NRQCD the quarkonium production cross section is organized as a double expansion in $\alpha_s$ and $v$. 

For this talk I will be concerned with $J/\psi$ production, though much of what I say below will be applicable to other $^3S_1$ quarkonium states, such as the $\psi(2S)$ and $\Upsilon(nS)$ states. For $J/\psi$ production, the leading matrix element is $\langle {\cal O}^{J/\psi}(^3S_1^{[1]})\rangle$  \footnote{Here $^{2S+1}L_J^{[i]}$ is the usual spectroscopic notation for the angular momentum  of the $c\bar{c}$ and $i=1,8$ denotes the color state.} which scales as $v^3$. Up to $v^4$ corrections this operator is equal to $|\psi_{c\bar{c}}(0)|^2$, where $\psi_{c\bar{c}}(0)$ is the wavefunction at the origin, which can be extracted from the width for $\Gamma[J/\psi \to e^+e^-]$ or from potential model calculations. Truncated to lowest order in $v$, NRQCD reproduces the color-singlet model of quarkonium production. The next most important operators for quarkonium production are $\langle {\cal O}^{J/\psi}(^3S_1^{[8]})\rangle $, $\langle {\cal O}^{J/\psi}(^1S_0^{[8]})\rangle$, and $\langle {\cal O}^{J/\psi}(^3P_J^{[8]})\rangle/m_c^2$ which scale as $v^7$. Unlike the leading color-singlet operator, the numerical values of these color-octet long-distance matrix elements (CO LDME)  must be extracted from fits to quarkonium production. 

A global analysis of the world's data on  quarkonium production with next-to-leading order (NLO) in $\alpha_s$ calculations of color-singlet and color-octet mechanisms is performed in Refs.~\cite{Butenschoen:2011yh,Butenschoen:2012qr}. For most observables, NRQCD at NLO does a reasonable job of reproducing  the data provided color-octet mechanisms are included. The extracted CO LDME are a factor of $\sim10^{-2}-10^{-3}$ smaller then the leading color-singlet matrix element, which is consistent with the NRQCD velocity scaling rules. The $\chi^2_{d.o.f.}$ for the global fit is 4.42 which is higher than one might hope for, but the authors of Refs.~\cite{Butenschoen:2011yh,Butenschoen:2012qr} argue that this is not unacceptable considering the  theoretical uncertainties in the NLO calculation are also significant. The most glaring problem for NRQCD is the {\it polarization puzzle}~\cite{Braaten:1999qk,Cho:1994ih,Beneke:1995yb}. Using the values of the CO LDME extracted in the fits of Refs.~\cite{Butenschoen:2011yh,Butenschoen:2012qr},
the polarization of the $J/\psi$ is predicted to  be transverse to its  momentum at high $p_T$ in collider experiments. Instead, all measurements of the  $J/\psi$ find essentially no polarization at all values of $p_T$~\cite{Affolder:2000nn,Abulencia:2007us,Chatrchyan:2013cla,Aaij:2013nlm}. 

Several groups have recently attempted to address the polarization puzzle by performing fits that focus exclusively on the high $p_T$ data from colliders. In Ref.~\cite{Chao:2012iv} a full NLO calculation including $^3P_J^{[8]}$ channels is fit to Tevatron data from CDF for $p_T > 7$ GeV. Ref.~\cite{Chao:2012iv} fits to both the $p_T$ spectra and the polarization and finds a rather wide range of CO LDME which can reproduce the $p_T$ spectra at both the Tevatron and the Large Hadron Collider (LHC) and give negligible polarization for all $p_T$. The extracted CO LDME are in disagreement with those extracted from the global fit in Refs.~\cite{Butenschoen:2011yh,Butenschoen:2012qr}. The lack of polarization for central  values of CO LDME obtained in the fit is attributed to a cancellation between contributions from $^3S_1^{[8]}$ and $^3P_J^{[8]}$ mechanisms, but Ref.~\cite{Chao:2012iv} also has fits for which production is dominated by the $^1S_0^{[8]}$ mechanism and $^3S_1^{[8]}$ and $^3P_J^{[8]}$ mechanisms are neglected. 

Ref.~\cite{Bodwin:2014gia} goes beyond NLO by working in the fragmentation approximation, in which the cross section for producing the $Q\bar{Q}$ is written as the convolution of cross sections for producing a parton $i$ with fragmentation functions for $i\to Q\bar{Q}$. DGLAP evolution of the fragmentation functions is  used to resum large logs of $p_T^2/(2m_c)^2$ in the single parton fragmentation contribution. Only color-octet mechanisms are included and the calculation is merged with the full NLO calculation for moderate $p_T$. Fitting to  $p_T$ spectra from both CDF and the LHC for $p_T > 10$ GeV, they find values of CO LDME which yield negligible polarization for all $p_T$, which is consistent with LHC data and greatly improves agreement with CDF data, especially at high $p_T$. Again the CO LDME extracted from the fit are inconsistent with the global fits in Refs.~\cite{Butenschoen:2011yh,Butenschoen:2012qr} and the $^1S_0^{[8]}$ LDME is significantly larger.

Recently an experimental group~\cite{Faccioli:2014cqa,Lourenco:2014sja} has done analyses of $\psi(2S)$ and $\Upsilon(3S)$ production that properly take into account correlations between the assumed polarization of the quarkonia and acceptances as a function of $p_T$. These analyses simultaneously fit $p_T$ spectra and polarization and include NLO calculations of $^3S_1^{[1]}$ mechanisms as well as $^1S_0^{[8]}$ and $^3S_1^{[8]}$ mechanisms. An interesting aspect of the study is how the extracted CO LDME depend on the lower limit on $p_T$ . They find that the  relative importance of $^1S_0^{[8]}$ versus $^3S_1^{[8]}$ mechanisms is very sensitive to the lowest value of the  $p_T$ included in the fit. If one includes only $\psi(2S)$ with $p_T > 7$ GeV, then fits prefer $^1S_0^{[8]}$ production mechanisms. When $\psi(2S)$ with lower $p_T$ are included, the fits  prefer $^3S_1^{[8]}$ mechanisms. Similarly, Ref.~\cite{Lourenco:2014sja} shows that for $p_T > 7$ GeV, $^1S_0^{[8]}$ mechanisms correctly predict the shape of the $J/\psi$ $p_T$ spectra, while lower $p_T$ data are more consistent with the $^3S_1^{[8]}$ production mechanism. They conclude that NRQCD factorization is only reliable for $p_T > 2 M$, where $M$ is the mass of the quarkonia, and that $^1S_0^{[8]}$ mechanisms dominate high $p_T$ quarkonium production.

\section{Heavy Quarkonium Fragmenting Jet Functions} 

It is clear that high $p_T$ data from colliders favors the $^1S_0^{[8]}$ production mechanism and this could explain the absence of any polarization of quarkonia at high $p_T$. The polarization puzzle can then be interpreted as a tension between high $p_T$ data from colliders with the rest of the world's data on $J/\psi$ production.  In light of this, it is important to find other observables at high $p_T$  that can be used to perform independent extractions of the CO LDME for $J/\psi$ production. At sufficiently high $p_T$, it is expected that single parton fragmentation should dominate particle production, then the $J/\psi$ should appear as part of a jet of particles produced by the initial high $p_T$ parton.
The main point of this talk is that cross sections for jets containing quarkonia provide alternative tests of NRQCD. The distribution of the quarkonium within the jet is sensitive to the underlying quarkonium production mechanism.

The cross section for a parton $i$ that fragments into a jet of energy $E$ and cone size $R$ with an identified hadron carrying a fraction $z$ of the jet's energy depends on a function called the fragmenting jet function (FJF), which was introduced recently in Ref.~\cite{Procura:2009vm} and further studied in Refs.~\cite{Liu:2010ng,Jain:2011xz,Jain:2011iu,Procura:2011aq,Jain:2012uq,Bauer:2013bza}. This function, which for a generic hadron $h$ will be denoted by ${\cal G}_i^h(E,R,\mu, z)$,  is closely related to the well-known fragmentation function that appears in inclusive hadron production~\cite{Collins:1981uw} and the jet functions that are well-known from the theory of jet cross sections~\cite{Ellis:2010rwa}.
The jet function for a jet with energy $E$ and cone size $R$ is denoted by $J_i(E,R, \mu)$, and is related to ${\cal G}_i^h(E,R,\mu, z)$ by 
\begin{equation}
J_i(E,R,\mu) =  \sum_h \int \frac{dz}{2(2\pi)^3} z \, {\cal G}_i^h(E,R,z,\mu) \, ,
\end{equation}
where $\mu$ is the renormalization scale.
This shows that the cross section for a jet with energy $E$ and cone size $R$ containing a hadron $h$ with energy fraction $z$ is obtained from the jet cross section
by replacing the function $J_i(E,R,\mu)$  with ${\cal G}_i^h(E,R,z,\mu)$. The other crucial property of  ${\cal G}_i^h(E,R,z,\mu)$ is its relation to the conventional fragmentation function, $D_j^h(z,\mu)$,
\begin{equation}\label{match}
{\cal G}_i^h(E,R,z,\mu) = \sum_j  \int_z^1 \frac{dz'}{z'} \,{\cal J}_{ij}(E,R,z',\mu) D^h_j(z/z',\mu) + O\left(\frac{\Lambda_{\rm QCD}^2}{4E^2\tan^2(R/2)}\right)\, ,
\end{equation}
where  ${\cal J}_{ij}(E,R,z',\mu)$ are calculable matching coefficients.  Large
logarithms in ${\cal J}_{ij}(E,R,z',\mu)$ are minimized at the jet energy scale $\mu_J = 2 E \tan(R/2)$. Thus, up to corrections that are order $\alpha_s(\mu_J)$, the ${\cal G}_i^h(E,R,z,\mu)$ is the fragmentation function evaluated at the jet energy scale $\mu_J$, $D^h_i(z,\mu_J)$.

\section{New Tests of NRQCD using Jet Observables}

In the 1990's Braaten and collaborators \cite{Braaten:1993mp,Braaten:1993rw,Braaten:1994vv,Braaten:1996jt} showed that heavy quarkonia fragmentation functions can be calculated in perturbation theory at the scale $2 m_Q$  in terms of NRQCD LDME. The $z$ dependence of the fragmentation function is calculable and the normalization of various NRQCD contributions is set by the LDME. To calculate the FJFs ${\cal G}_i^h(E,R,z,\mu)$ when $h$ is a quarkonium, one simply needs to evolve the NRQCD fragmentation functions from the scale $2m_Q$ to the scale $\mu_J$ using DGLAP evolution and then perform the convolution with the matching coefficients in Eq.~(\ref{match})~\cite{Baumgart:2014upa}.

At the scale $2 m_c$, the different NRQCD production mechanisms give rise to very different $z$ dependence in the $g\to J/\psi$ fragmentation function. For example,  gluon fragmentation to quarkonium in the $^3S_1^{[8]}$ channel is proportional to $\delta(1-z)$ at lowest order in $\alpha_s(2 m_Q)$ because a gluon can turn into a $^3S_1^{[8]}$ $Q\bar{Q}$ pair without emitting any gluons. Two gluons must be emitted in order for a fragmenting gluon to transition to a  $Q\bar{Q}$ pair in a $^3S_1^{[1]}$ state, thus the $^3S_1^{[1]}$ contribution to the gluon fragmentation function is a rather broad distribution in $z$. The $^1S_0^{[8]}$ and $^3P_J^{[8]}$ contributions to the gluon fragmentation also have distinct $z$ dependence. Fig.~\ref{HQQFF} shows the various NRQCD contributions to the $g\to J/\psi$ fragmentation function as well as the color-singlet $c\to J/\psi$ fragmenation function. The normalization of each distribution is arbitrary. As stated earlier, the $^3S_1^{[8]}$ contribution is proportional to $\delta(1-z)$ and is represented by a vertical line at $z =1$. 

\FIGURE[h]{
\includegraphics[width = 2.5 in, angle=270]{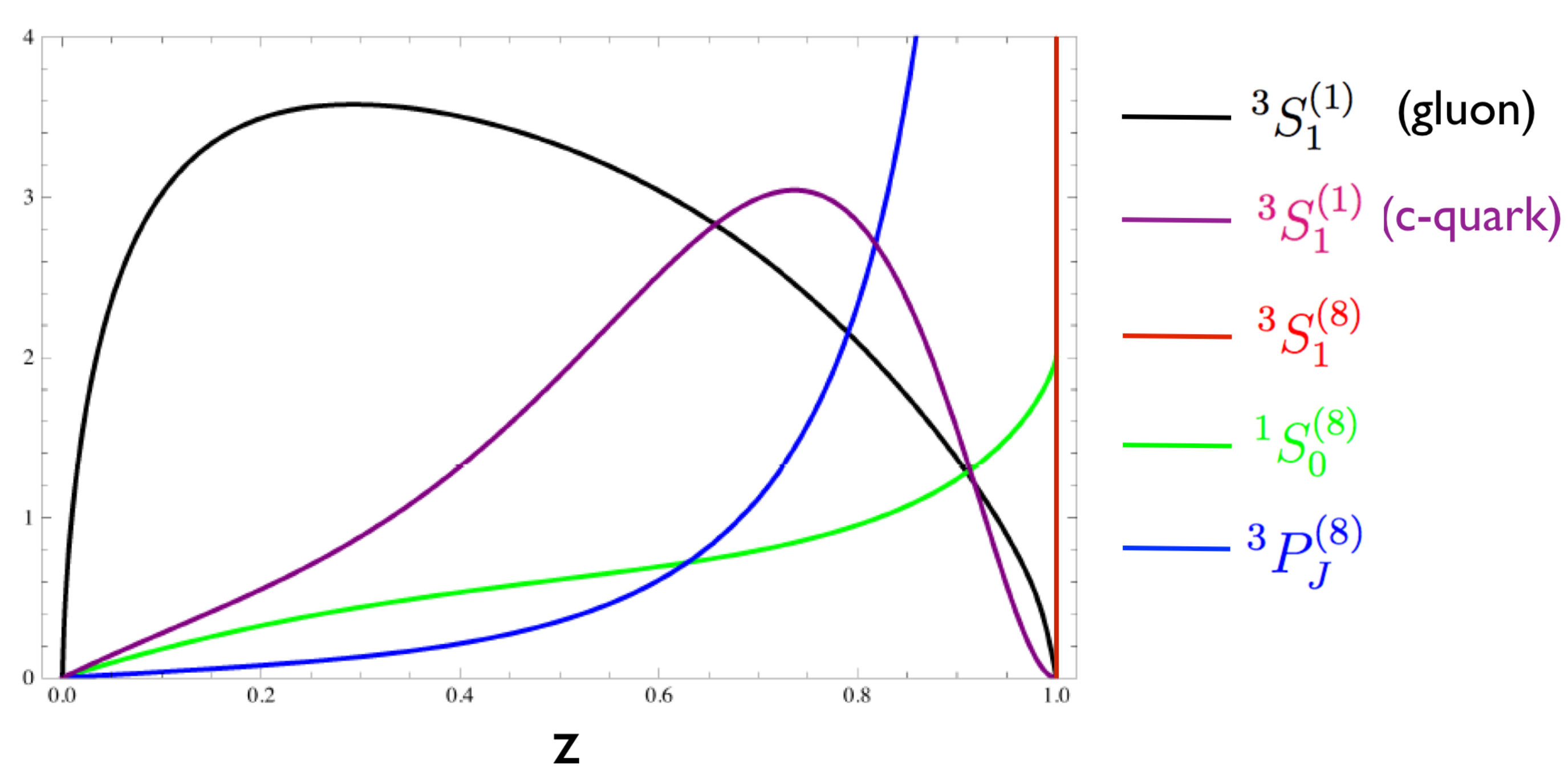}

{ \caption[1] {Leading order NRQCD predictions for the $J/\psi$ fragmentation functions for gluon and charm quark fragmentation functions at $\mu=2m_c$. }
\label{HQQFF}}
}

\begin{figure}[!t]
\begin{center}
\includegraphics[width=15cm]{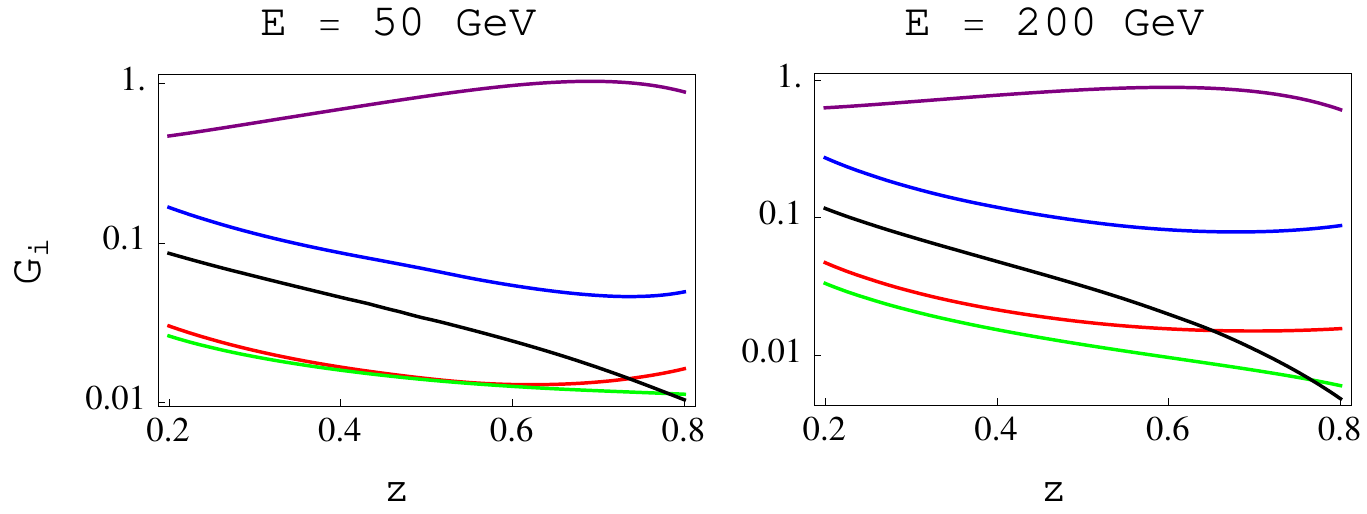}
\end{center}
\vspace{- 0.5 cm}
\caption{
\label{logFJF}
\baselineskip 3.0ex
Different NRQCD contribution to the gluon FJF and the  $^3S_1^{[1]}$ charm quark  FJF~\cite{Baumgart:2014upa}. }
\end{figure}

\begin{figure}[!h]
\begin{center}
\includegraphics[width=15cm]{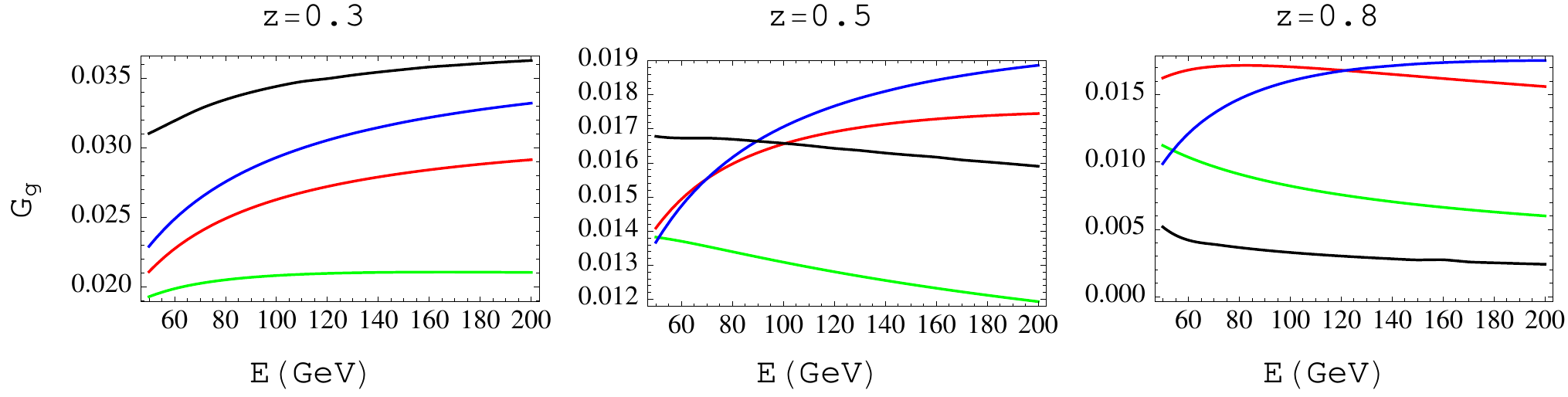}
\end{center}
\vspace{- 0.5 cm}
\caption{
\label{fixedz} 
\baselineskip 3.0ex
The energy dependence of the four different contributions to the  gluon FJF for fixed $z=0.3$, $0.5$, and $0.8$.  in this plot I have scaled the $^3P_J^{[8]}$ function down by a factor of 5 and $^3S_1^{[1]}$ down by 2~\cite{Baumgart:2014upa}. }
\end{figure}

\begin{figure}[!t]
\begin{center}
\includegraphics[width=15cm]{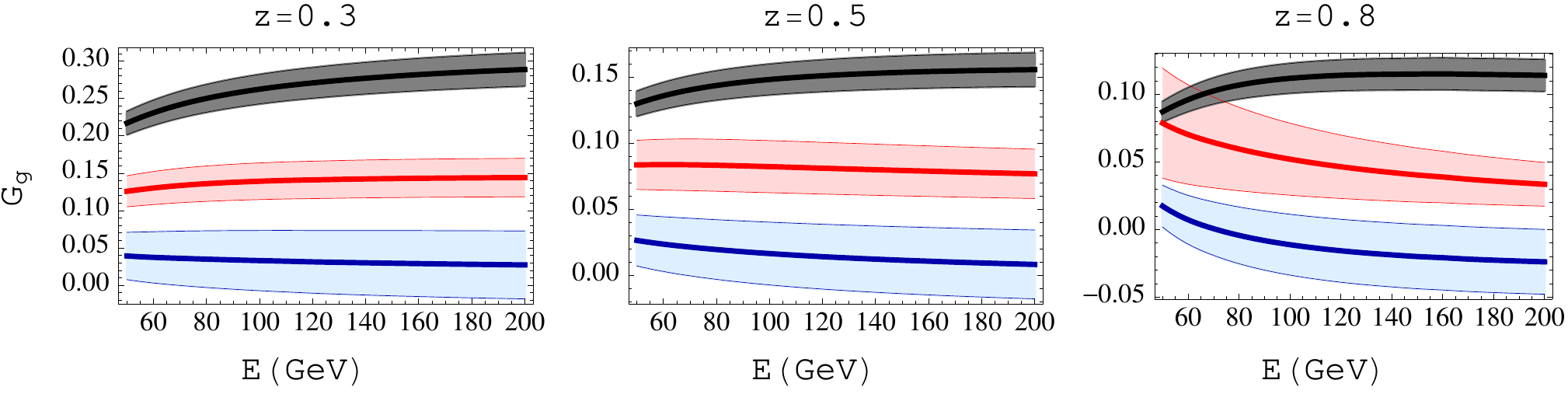}
\end{center}
\vspace{- 0.5 cm}
\caption{
\label{zcompare} 
\baselineskip 3.0ex
The gluon FJF at fixed momentum fraction for the LDME extracted in Refs.~\cite{Butenschoen:2011yh,Butenschoen:2012qr} (gray),
Ref.~\cite{Chao:2012iv} (blue), and Ref.~\cite{Bodwin:2014gia} (red), from Ref.~\cite{Baumgart:2014upa}. }
\end{figure}

The evolution of the heavy quarkonium fragmentation function from the scale $2 m_Q$ up to the scale $\mu_J$ significantly modifies the shape of these $z$ distributions,  but the various mechanisms still can be differentiated. The FJF ${\cal G}_i^{J/\psi}(E,R,z,\mu)$ is a function of  $E$, $R$, and $z$, and  the dependence on $E$ and $R$ is entirely due to the dependence on the scale $\mu_J$.  Ref.~\cite{Baumgart:2014upa} calculates the FJFs with $R=0.4$ and varies $E$ and $z$. 
Fig.~\ref{logFJF} shows the different NRQCD contributions to the gluon FJF as well as the color-singlet charm quark FJF as a function of $z$ for different values of $E$. The  color coding is the same as in Figure~\ref{HQQFF}. The LDME used in these calculations are the central values obtained from the global fit in Refs.~\cite{Butenschoen:2011yh,Butenschoen:2012qr}. The various NRQCD production mechanisms yield different $z$ dependence though the differences are not nearly as dramatic as  in Fig.~\ref{HQQFF}.   
The different contributions to the gluon FJF as a function of $E$ for fixed values of $z$ are shown in Fig.~\ref{fixedz}~\cite{Baumgart:2014upa}. Note that
the $^1S_0^{[8]}$ contribution to the gluon FJF has negative slope in $E$ for $z> 0.5$, which is qualitatively different from the other color-octet mechanisms. 
Observing this feature in jets containing $J/\psi$ would be a nontrivial test of the hypothesis that the $^1S_0^{[8]}$ channel dominates high $p_T$ production of $J/\psi$.
\begin{figure}[!t]
\begin{center}
\includegraphics[width=15cm]{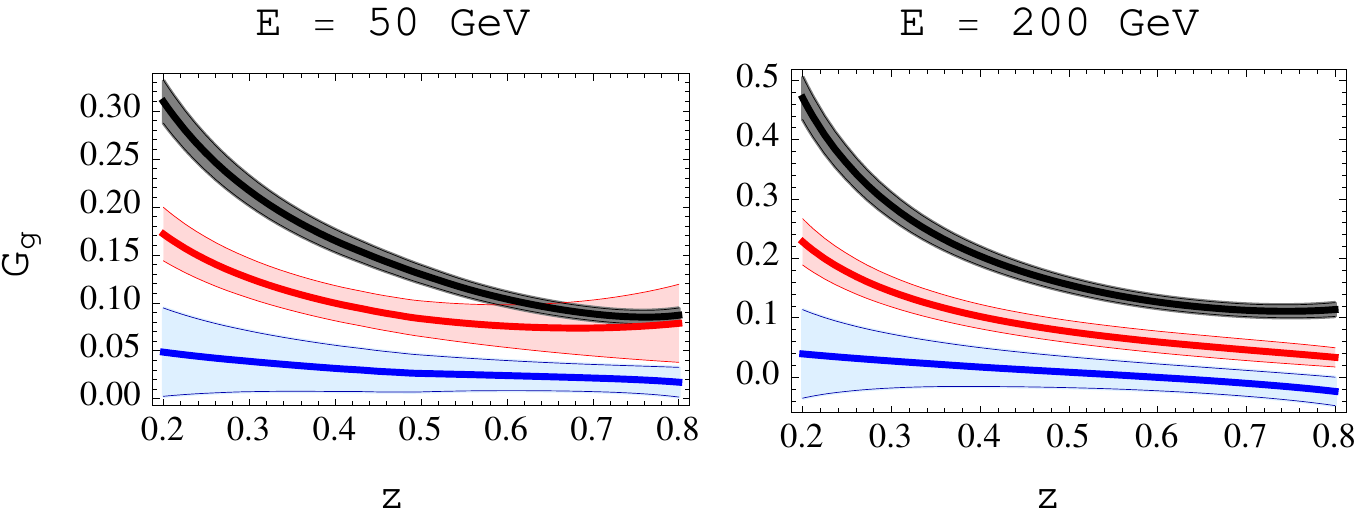}
\end{center}
\vspace{- 0.5 cm}
\caption{
\label{fit_compare} 
\baselineskip 3.0ex
The gluon FJF at fixed energy for the LDME extracted in Refs.~\cite{Butenschoen:2011yh,Butenschoen:2012qr} (gray),
Ref.~\cite{Chao:2012iv} (blue), and Ref.~\cite{Bodwin:2014gia} (red), from Ref.~\cite{Baumgart:2014upa}.}
\end{figure}
 At the LHC $J/\psi$ production from  gluon fragmentation will dominate over charm quark fragmentation. Then measurements of high $p_T$ $J/\psi$ within jets should be sensitive to  ${\cal G}_g^{J/\psi}(E,R,z,\mu_J)$. To see if measurement of this FJF can discriminate between various extractions of the CO LDME, ${\cal G}_g^{J/\psi}(E,R,z,\mu_J)$ is computed for three different extracted values of the CO LDME appearing in the literature~\cite{Butenschoen:2011yh,Butenschoen:2012qr,Chao:2012iv,Bodwin:2014gia}. The results are plotted in Fig.~\ref{fit_compare} as a function of $z$ for fixed values of $E$, and in Fig.~\ref{zcompare} as a function of $E$ for fixed values of $z$~\cite{Baumgart:2014upa}. The resulting FJFs are different even taking into account uncertaintites in the matrix elements. For the values of CO LDME extracted  in global fits to $J/\psi$ production there is a qualitative difference in the slope as a function of $E$ for $z>0.5$  compared to the values of the CO LDME from the other two groups who fit to only to high $p_T$ data from colliders. This behavior is expected as these fits yield larger values for $\langle {\cal O}^{J/\psi} ({}^1S^{[8]}_0)\rangle$ than the global fits. 

\section{Cross sections for $e^+e^-$, $pp$ collisions}

For $e^+e^-$ collisions, a general framework for calculating an N-jet cross section was developed in Ref.~\cite{Ellis:2010rwa}. The cross section takes the schematic form 
\begin{equation}\label{master}
d\sigma = H_{i1,i2,...,iN} \times J_{i1} \otimes J_{i2} \otimes ... \otimes J_{iN} \otimes S_{i1,i2,...,iN} \, .
\end{equation}
Here $H_{i1,i2,...,iN} $ is the hard function which contains the 
large virtual corrections to the cross section. The functions $J_{in}$ are the jet functions and describe radiation collinear to one of the jets.
The soft function is denoted $S_{i1,i2,...,iN}$ and describes soft radiation connecting the jets. A sum over partons  is implied in Eq.~(\ref{master}). Ref.~\cite{Ellis:2010rwa} calculated cross sections for the jet substructure variable called angularity~\cite{Berger:2003iw}, defined by  $\tau_a = \frac{1}{\omega} \sum_i (p_i^+)^{1-a/2} (p_i^-)^{a/2}$, $\omega = \sum_i p_i^-$. Here for both $\tau_a$ and $\omega$ the sum is over the hadrons in the jet, and $p_i^\pm$ are light-cone components of the hadron momenta. In the terminology of Ref.~\cite{Ellis:2010rwa}, a jet is called unmeasured if only its total $E$ is known, and measured if the substructure $\tau_a$ is also observed. Unmeasured jet functions enter the cross section multiplicatively, while for measured jets $\otimes$ represents a convolution over $\tau_a$ which includes both measured jets and the soft function.

\begin{figure}[!t]
\begin{center}
\includegraphics[width=15cm]{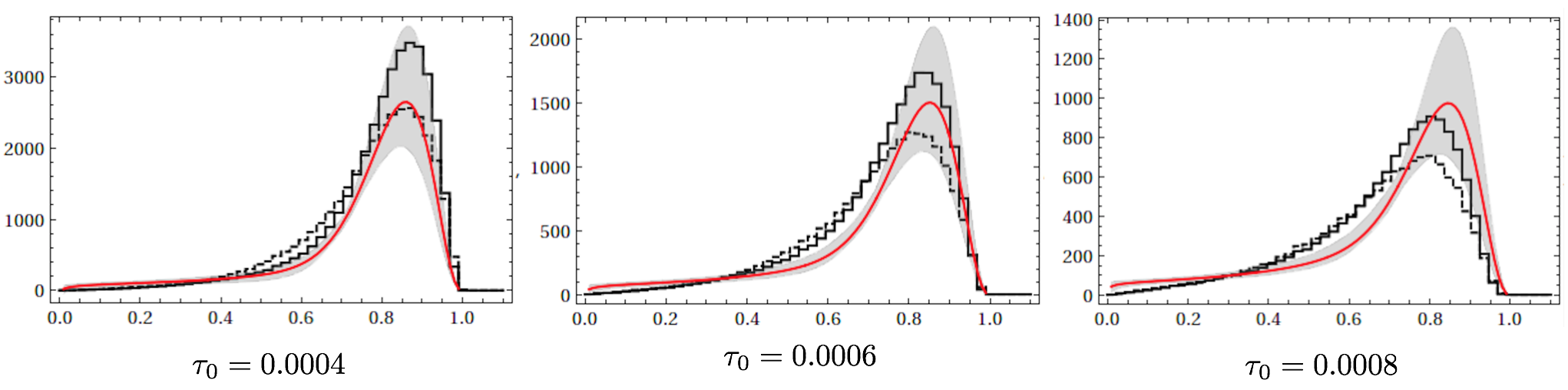}
\end{center}
\vspace{- 0.5 cm}
\caption{
\label{Bmeson_z} 
\baselineskip 3.0ex
The $z$ distribution for fixed $\tau_0$ for $B$ mesons in dijet events in $e^+e^-$ at $\sqrt{s}=600\,{\rm GeV}$.
Red lines is NLL prediction, gray band is theoretical uncertainty, solid (dotted) line is prediction from Pythia (Herwig) event generator. }
\end{figure}

\begin{figure}[!t]
\begin{center}
\includegraphics[width=15cm]{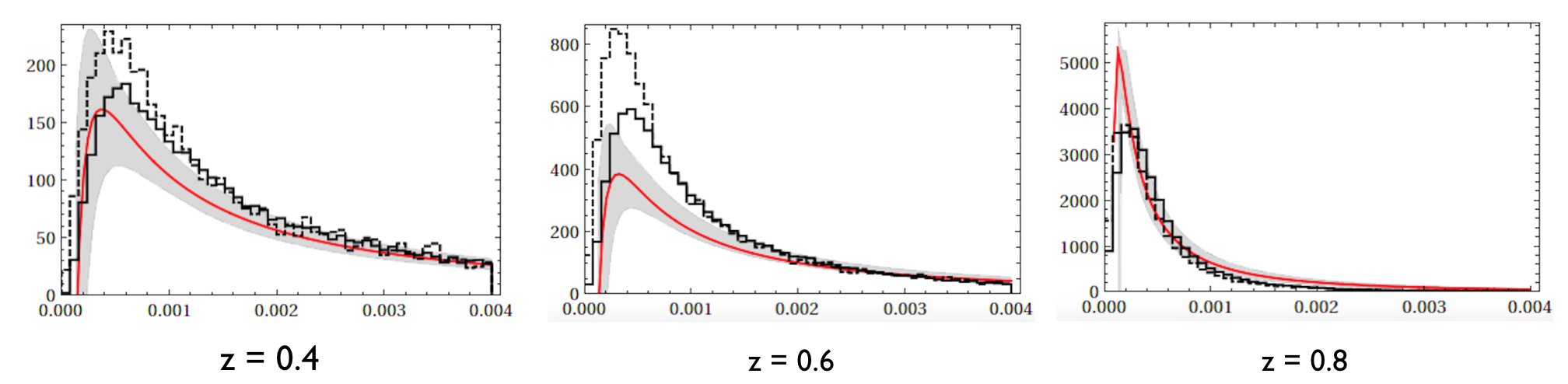}
\end{center}
\vspace{- 0.5 cm}
\caption{
\label{Bmeson_tau} 
\baselineskip 3.0ex
The $\tau_0$ distribution for fixed $z$ for $B$ mesons in dijet events in $e^+e^-$ at $\sqrt{s}=600\,{\rm GeV}$.
Red lines is NLL prediction, gray band is theoretical uncertainty, solid (dotted) line is prediction from Pythia (Herwig) event generator. }
\end{figure}

In order to calculate cross sections for jets containing identified hadrons at an $e^+e^-$ collider, one simply replaces one of the $J_{in}$ in Eq.~\ref{master} with the appropriate FJF.
By the evolving the hard, jet and soft functions to their appropriate scales using the leading order renormalization group equation (RGE), one obtains analytic next-to-leading log (NLL) resummed sections. Since the ${\cal G}_i^h(E,R,z, \mu)$ obeys the same RGE as $J_i(E,R,\mu)$, the expression for the resummed cross section is essentially the same as found in Ref.~\cite{Ellis:2010rwa}. To test this formalism for calculating the distribution of hadrons in jets, we consider $e^+e^- \to 2\, {\rm jets}$, where a $B$ meson is observed in one of the jets~\cite{BDHLMM}. In our study the angularity of the jet containing the $B$ meson is also measured, so the corresponding FJF requires a different set of angularity dependent matching coefficients for an equation analogous to Eq.~\ref{match}~\cite{BDHLMM}. For the fragmentation function at the scale $\mu = m_b$, we use the $B$ meson fragmentation extracted from LEP data  in Ref.~\cite{Kniehl:2008zza}. The differential cross as a function of $z$ for various values of $\tau_0$ is shown in Fig.~\ref{Bmeson_z}, and as function of $\tau_0$ for various values of $z$ is shown in Fig.~\ref{Bmeson_tau}. Because we are interested in jets whose energy is comparable to those at the LHC, we choose $\sqrt{s}= 600$ GeV. The solid redline is the prediction at NLL and the gray band corresponds to the theoretical uncertainty obtained by varying hard, jet, and soft scales in the factorization theorem by a factor of two.  Also shown are the same distributions calculated with Monte Carlo generators Pythia (solid lines) and Herwig (dotted lines). We see that within uncertainties our NLL calculations agree well with Monte Carlo for a large range of $z$ and $\tau_0$ giving us confidence that our NLL calculation with the FJFs correctly calculates both the jet substructure and the distribution of the $B$ mesons within the jet.

It is possible to test the NLL resummed cross sections against Monte Carlo predictions for quarkonia production in $e^+e^-$ collisions. Since we expect gluon fragmentation to be the dominant mechanism of production at the LHC, we focus on three jet events in which the $J/\psi$ appears in the jet initiated by the gluon. This is not a physical observable as one must also include $q,\bar{q}\to J/\psi$ fragmentation, however, this is expected to be a smaller contribution. We compare the NLL calculation with Monte Carlo generated events in which the $J/\psi$ is required to appear in the jet initiated by the gluon. Our studies have shown that while the 
$\tau_0$ distributions are in agreement with the NLL calculation, the $z$ distribution for the $J/\psi$ is not in agreement~\cite{BDHLMM}. The agreement with $\tau_0$ distributions reflects the fact that $\tau_0$ is a global observable which depends on all the particles in the jet, and is insensitive to the behavior of a single hadron in that jet.  Pythia  gives identical $z$ distributions for all color-octet mechanisms, which is inconsistent with the analytic studies using the FJFs. We believe this is because Pythia uses a simplified model of color-octet $Q\bar{Q}$ fragmentation that is not consistent with NRQCD predictions for the $g\to Q\bar{Q}$ fragmentation functions. For plots of the $z$ and $\tau_0$ distributions we refer the reader to publicly available conference proceedings~\cite{conf}. Further work is needed to make existing Monte Carlo generators consistent with NRQCD predictions for quarkonia production within jets. 

Finally, in order to make analytic NLL predictions for the production of heavy mesons within jets at the LHC, one must generalize the factorization formulae of Eq.~\ref{master}
to a form suitable for $pp$ collisions.\footnote{An alternative approach is to study jet fragmentation by constructing ratios of the differential cross sections for jets with identitified hadrons with the inclusive jet cross section. The observable is the ratio of the FJF to the jet function. This was successfully compared to LHC data on light meson and  $D$ meson production in Ref.~\cite{Chien:2015ctp}.} Recently, substantial progress in this direction has been made in Ref.~\cite{Hornig:2016ahz}. First, Eq.~\ref{master} must be generalized to include parton distribution functions for the incoming beams. In addition, observables in hadron collisions are defined to be invariant under boosts along the beam axis. In $e^+e^-$ collisions, a cut is placed on the energy of soft radiation outside the jets. In $pp$ collisions, the constraint on energy outside the jets is a cut on $p_T$ for radiation with a certain rapidity range. Ref.~\cite{Hornig:2016ahz} defines a generalization of $\tau_a$ that is invariant under boosts along the beam axis and shows that both unmeasured and unmeasured jet functions can be obtained from the corresponding jet functions in $e^+e^-$ by a simple rescaling of their arguments. A new soft function is required for these cross sections and is calculated in 
Ref.~\cite{Hornig:2016ahz} for dijet events. With the developments of Ref.~\cite{Hornig:2016ahz} and Ref.~\cite{Baumgart:2014upa}, the theoretical tools are now available to provide analytic calculation of resummed cross sections for jets with quarkonia that will allow for novel tests of NRQCD. 

\acknowledgments
I would like to acknowledge my collaborators: R. Bain, M. Baumgart,  L. Dai, A. Hornig, Y. Makris, A. Leibovich, and I. Rothstein. I am supported in part by the Director, Office of Science, Office of Nuclear Physics, of the U.S. Department of Energy under grant numbers DE-FG02-05ER41368. I also wish to  acknowledge the hospitality of Duke-Kunshan University and the theory groups at Brookhaven National Laboratory, Los Alamos National Laboratory, and UC Irvine during the completion of this work.

\end{document}